# Giant piezomagnetic response driven by spin reorientation in the topological Kagome magnet $TbMn_6Sn_6$

Hanyang Qian[1, 2], Wen Sun[1, 2], Juntao Huo[1, 2], Guowei Li[1, 2]

1. Ningbo Institute of Materials Technology and Engineering, Chinese Academy of Sciences, Ningbo 315201, China.

2. Center of Materials Science and Optoelectronics Engineering, University of Chinese Academy of Sciences, 19 A Yuquan Rd, Shijingshan District, Beijing 100049, China.

Email: liguowei@nimte.ac.cn.

**Abstract**

Achieving a large strain response to a small magnetic-field variation remains a challenge in magnetomechanical materials, particularly in topological systems where the coupling between spin reconstruction and lattice deformation remains unclear. Here, we report an anomalous magnetostrictive response associated with field-induced spin reorientation in the topological Kagome magnet $TbMn_6Sn_6$. Below its spin-reorientation temperature of 310 K, an in-plane magnetic field induces an abrupt increase in magnetostriction, which exceeds 120 ppm at 240 K. This transition produces a giant differential piezomagnetic coefficient of approximately 5000 ppm $T^{-1}$, far exceeding the values reported for other topological magnetic materials and commercial FeGa alloys, and placing $TbMn_6Sn_6$ among the most field-sensitive magnetostrictive materials. The field position of this peak agrees with the anomalies in magnetization and magnetotransport, demonstrating that the large differential response is tied to field-induced spin reorientation. Thermal expansion also changes across the spin-reorientation boundary. Raman measurements reveal corresponding anomalies in phonon energy and linewidth, indicating that the reorientation modifies the lattice dynamics. The combined results support a mechanism in which competing magnetic anisotropies of the Tb and Mn sublattices, coupled through Tb-Mn exchange, create magnetic configurations with different equilibrium strains. This finding establishes spin reorientation as a route to giant piezomagnetic responses in topological magnets and provides a basis for magnetomechanical sensing and switching near room temperature.

**Introduction**

Magnetostriction, the deformation of a solid in response to a magnetic field,

reflects the coupling between magnetic order and lattice [1]. It is technologically attractive, because it enables the conversion of magnetic stimuli to mechanical responses and vice versa [2]. The performance of a magnetostrictive material is not determined by its maximum strain alone, but also by its differential strain response to magnetic-field variations, operating field, hysteresis, and reversibility [3, 4]. Conventional materials such as Terfenol-D and Galfenol illustrate different trade-offs between achievable strain and mechanical properties [5-7]. These considerations motivate the search for alternative mechanisms that can produce a pronounced and controllable strain response under magnetic fields. In particular, magnetic-field-induced transitions between competing magnetic configurations, such as the spin reorientation transition (SRT) [8-10], offer a potential route to strongly field-dependent magnetoelastic responses, because changes in spin orientation and magnetic order can modify the equilibrium lattice strain.

Recently, magnetic topological materials have provided an additional platform for exploring this problem. For certain magnetic topological materials, the interplay among magnetic order, spin-orbit coupling and crystal symmetry determines the electronic band crossings, gaps and Berry curvature. Consequently, changes in magnetic configuration may modify both the magnetoelastic response and topology-related electronic properties [11-16], creating opportunities to investigate and potentially control magnetic, electronic and lattice responses within such material system. This possibility also motivates exploring magnetic topological materials for multifunctional magnetomechanical sensing and actuation. An illustrative example is the Kagome antiferromagnet $Mn_3Sn$, which exhibits pronounced, nearly linear and non-saturating magnetostriction [17, 18]. This response has been attributed to field-induced twisting of the in-plane spin texture and its coupling to the lattice. This finding indicates that field-driven evolution of a non-collinear spin configuration can generate substantial and strongly anisotropic strain. It also raises a question: can a more pronounced magnetic reconstruction, such as field-induced spin reorientation, provide another route to generating a distinct magnetostrictive response in a magnetic topological material?

To address these questions, the in-plane magnetostriction of single-crystalline $TbMn_6Sn_6$, a layered Kagome ferromagnet belonging to $RMn_6Sn_6$ (R = rare-earth element) family [19], has been systematically investigated. In this compound, the orientation of the magnetic moments in the Mn Kagome layers is relevant to the formation of the topology-related Chern gap, providing a specific connection between magnetic configuration and electronic structure [20-22]. $TbMn_6Sn_6$ undergoes a well-established SRT at ~310 K ($T_{SR}$), with the preferred moment orientation changing from the *ab* plane towards the *c* axis upon cooling. Below $T_{SR}$, an applied in-plane magnetic field can drive a further reversed magnetic reorientation from the *c* axis to the *ab* plane [23]. This combination of a well-defined thermal SRT and a field-tunable magnetic response makes $TbMn_6Sn_6$ a suitable system for investigating the relationship between magnetic reconstruction and magnetostriction, as well as the associated electronic and lattice dynamic behaviors.

Here, we investigate the field-induced magnetostrictive response of $TbMn_6Sn_6$ as a function of temperature and magnetic field, together with magnetization, magneto-transport, thermal-expansion, and temperature-dependent Raman measurements. The magnetostriction exhibits an abrupt enhancement in the field-induced spin-reorientation regime, with its critical field corresponding to anomalies in magnetization and magneto-transport. Thermal expansion provides an additional measure of the lattice response across the thermal SRT, while Raman spectroscopy reveals anomalies in the temperature evolution of the in-plane $E_g$ mode phonon frequency and linewidth near the SRT. These measurements connect the macroscopic magnetoelastic response with the lattice-dynamical sensitivity of the material to magnetic reconfiguration. We discuss the results in terms of competing Tb- and Mn-sublattice anisotropies, Tb-Mn exchange, and spin-orientation-dependent magnetoelastic coupling. The study thus provides a combined characterization of field-induced magnetostriction and phonon response in $TbMn_6Sn_6$, offering further experimental insight into the interplay between magnetic reconstruction and lattice dynamics in a Kagome ferromagnet.

**Experimental methods**

Single crystal of $TbMn_6Sn_6$ were synthesized by self-flux method. High purity Tb, Mn and Sn were taken in a stoichiometric molar ratio of 1: 6: 20. The constituent elements were placed in alumina crucible and sealed in an evacuated quartz tube. The stoichiometric samples were heated to 1473 K over 20 h and kept there for 24 h before slowly cooled to 873 K over 150 h. The tube was taken out of the furnace at 873 K and decanted with a centrifuge to separate $TbMn_6Sn_6$ crystals from the flux. The single-crystal XRD patterns were recorded using a Bruker D8 Advance Davince X-ray diffractometer with Cu $K_\alpha$ radiation at ambient temperature. The morphology and compositions were determined by using scanning electron microscopy (SEM; FEI Quanta FEG 250) coupled with energy-dispersive X-ray spectroscopy (EDS). SAED patterns and STEM images were collected by aberration-corrected high-angle annular dark-field scanning transmission electron microscopy (Spectra 300 S/TEM). A superconducting quantum interference device (SQUID) magnetometer (MPMS 3, Quantum Design) was used to measure magnetic properties. The magnetostriction was measured by using strain gauge through a Physical Property Measurement System (PPMS-DynaCool, Quantum Design). The electrical and transport measurement were also conducted by using PPMS. The LabRAM Odyssey (Horiba) was deployed to conducted temperature-dependent Raman spectroscopy measurement.

**Results and discussion**

The as-grown $TbMn_6Sn_6$ single crystals exhibit a shiny metallic surface with well-defined facets, characteristic of layered crystal morphology. The crystal orientation was first examined by X-ray diffraction (XRD). As shown in Fig. 1(a), only a series of sharp (001) diffraction peaks are observed, demonstrating that the naturally exposed surface corresponds to the (001) plane and confirming the excellent out-of-plane crystallographic orientation of the single crystal. The chemical composition was further examined by SEM equipped with EDS. The elemental analysis yields an average composition of $TbMn_{6.57}Sn_{6.84}$, which is close to the nominal stoichiometry and shows no detectable compositional inhomogeneity over the scanned area, indicating good chemical uniformity of the grown crystals (Fig. 1(b)).

To further resolve the atomic structure, HAADF-STEM was performed along the [001] zone axis (Fig. 1(c)). As shown in Fig. 1(d), the atomic-resolution image clearly resolves the characteristic Kagome framework of the Mn sublattice. Owing to the Z-contrast imaging mechanism of HAADF-STEM, atomic columns with different atomic numbers can be readily distinguished. In the [001] projection, the brightest atomic columns are assigned to Tb atoms based on the Z-contrast and the crystallographic model, while the Mn atoms form the characteristic Kagome network. The remaining atomic columns can be assigned to Sn atoms according to the crystallographic structure model. The experimentally observed atomic arrangement agrees well with the reported crystal structure of $TbMn_6Sn_6$, confirming the successful growth of the layered Kagome phase [24]. These results demonstrate the high structural quality and chemical homogeneity of the as-grown $TbMn_6Sn_6$ single crystals, providing a solid materials basis for the subsequent investigations.

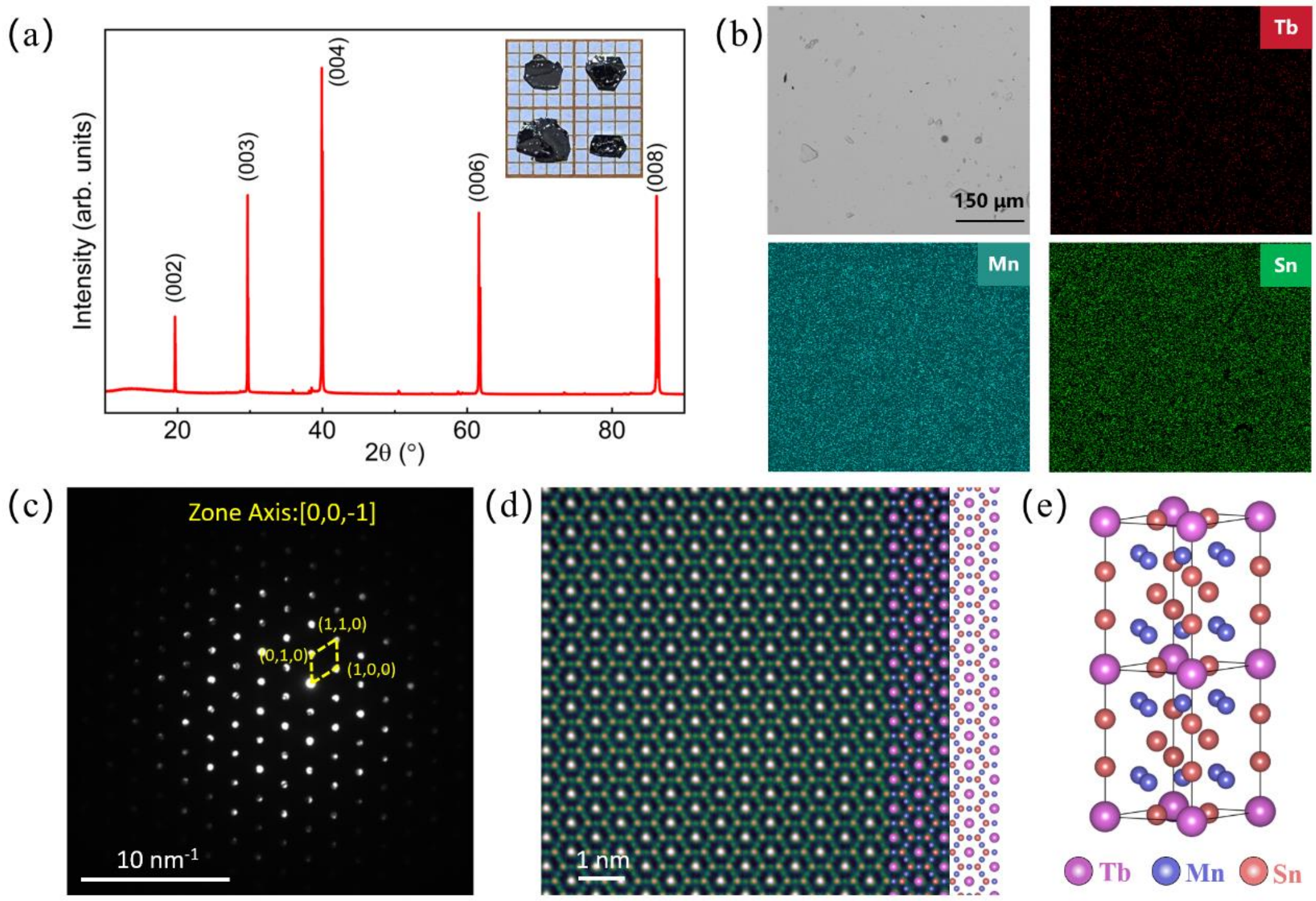

Figure 1. Characterization of $TbMn_6Sn_6$ single crystal. (a) XRD pattern, the inset shows the morphology of the as-grown $TbMn_6Sn_6$ single crystals. (b) EDS mappings of $TbMn_6Sn_6$. (c) SAED patterns. (d) Z-contrast resolution STEM image viewed along the c axis. (e) Atomic models of $TbMn_6Sn_6$ lattice.

Thermomagnetic curves (*M*-*T*) with field respectively applied along in-plane and out-of-plane directions for $TbMn_6Sn_6$ were measured upon zero-field-cooling (ZFC) and field-cooling (FC) process (Fig. 2 (a, c)). A pronounced magnetic anomaly is observed around 310K for both field directions, corresponding to the well-known spin-reorientation ($T_{SR}$). This result agrees well with previous reports that easy magnetization axis switches from the in-plane above $T_{SR}$ to the *c* axis below $T_{SR}$ [22]. The magnetic-field-dependent magnetization (*M*-*H*) were also investigated for both field directions at temperatures across $T_{SR}$. For $H \perp c$, step-like increase in magnetization on *M*-*H* curves was observed below $T_{SR}$. Such a step-like increase is the characteristic of a field-induced spin-reorientation from the c-axis-dominated state toward the in-plane configuration. It is seen that the critical magnetic field needed to reorient the spin direction increases upon lowering temperature. Additionally, the in-plane *M*-*T* curved recorded under various magnetic field, as seen in the inset of Fig. 2(b), show that the in-plane magnetic field could lower the $T_{SR}$. These results provide direct evidence that the spin reorientation is governed by the competition between Zeeman energy and magnetocrystalline anisotropy, which also establishes the magnetic basis for the following magnetostriction measurements. A similar field-induced magnetization anomaly is observed for $H//c$ above $T_{SR}$, indicating that the magnetic-field-induced spin-reorientation remains tunable by external magnetic field on both sides of $T_{SR}$ despite the distinct easy-axis anisotropy (Fig. 2(d)). The well-defined magnetic-field-induced spin-reorientation provides an ideal opportunity to directly correlate spin reconstruction with lattice deformation.

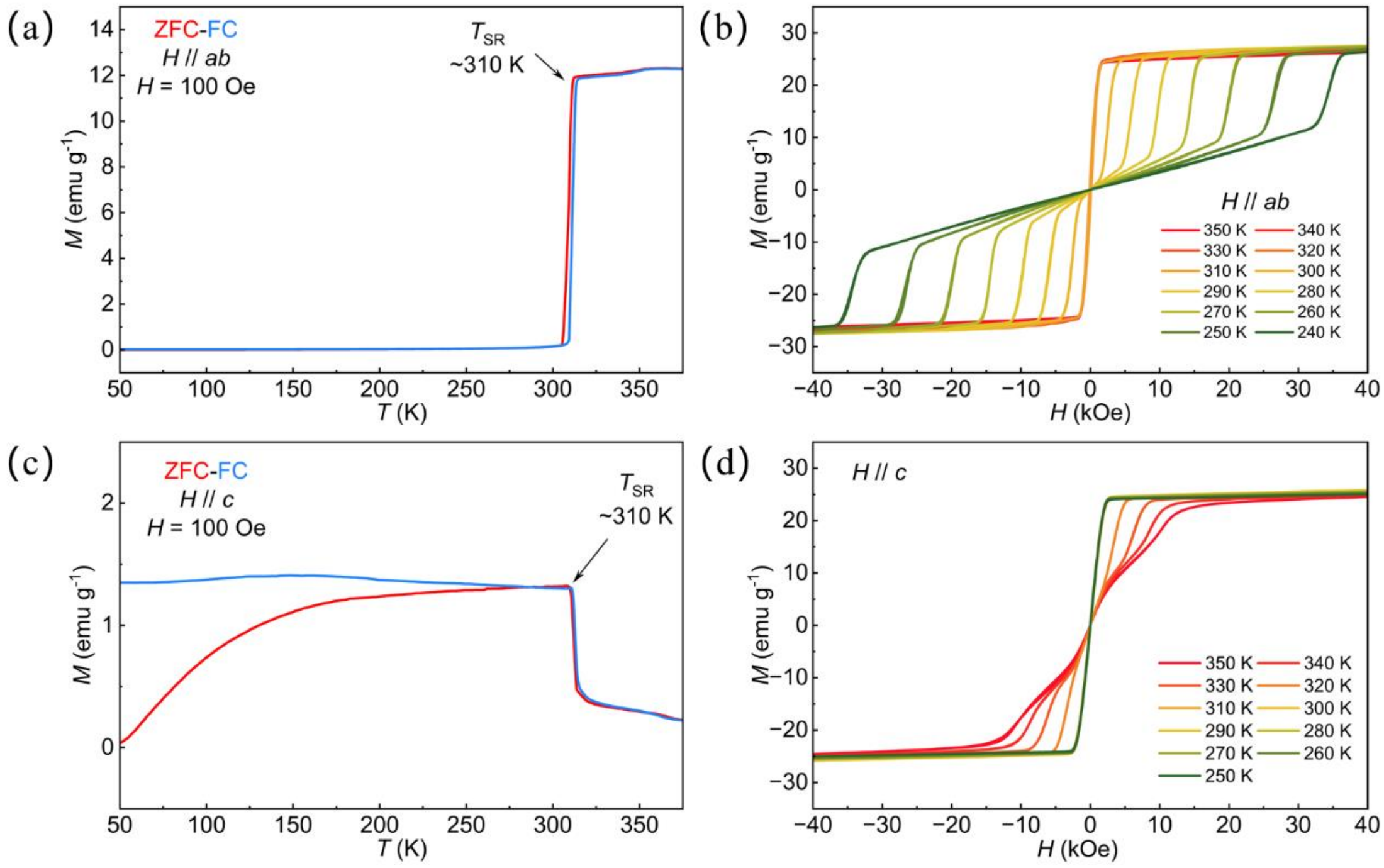

Figure 2. (a, c) Thermomagnetic curves for $TbMn_6Sn_6$ upon zero-field-cooling and field-cooling process measured along the ab-plane and c-axis. (b, d) Magnetic-field dependent magnetization measured along the ab-plane and c-axis at various temperature across $T_{SR}$.

Considering that the spin reorientation is induced by an in-plane magnetic field below $T_{SR}$ and $TbMn_6Sn_6$ possesses a quasi-two-dimensional layered crystal structure, the magnetostriction was measured using a strain gauge attached along the in-plane direction of $TbMn_6Sn_6$ single crystal with magnetic field applied perpendicular to the $c$ axis (parallel to the in-plane direction). This configuration allows the in-plane lattice deformation to be probed directly.

Fig. 3 presents the field-dependent in-plane magnetostriction measured at various temperature across $T_{SR}$. Above the $T_{SR}$, $TbMn_6Sn_6$ exhibits a negative, nearly linear and non-saturating magnetostrictive response. Such behavior resembles that in $Mn_3Sn$. Although the microscopic origin of the linear magnetostriction in $TbMn_6Sn_6$ remains to be clarified, the similar phenomenology suggests that appreciable magnetoelastic coupling may generally exist in magnetic topological materials.

A striking evolution of the magnetostriction emerges upon cooling below $T_{SR}$. As seen in the Fig. 3(a), the initial magnetostrictive response changes from negative to positive. More importantly, an abrupt increase in magnetostriction occurs once the applied magnetic field exceeds a critical value, followed by a rapid saturation at higher fields, indicating a qualitatively different field-dependent lattice response below $T_{SR}$.

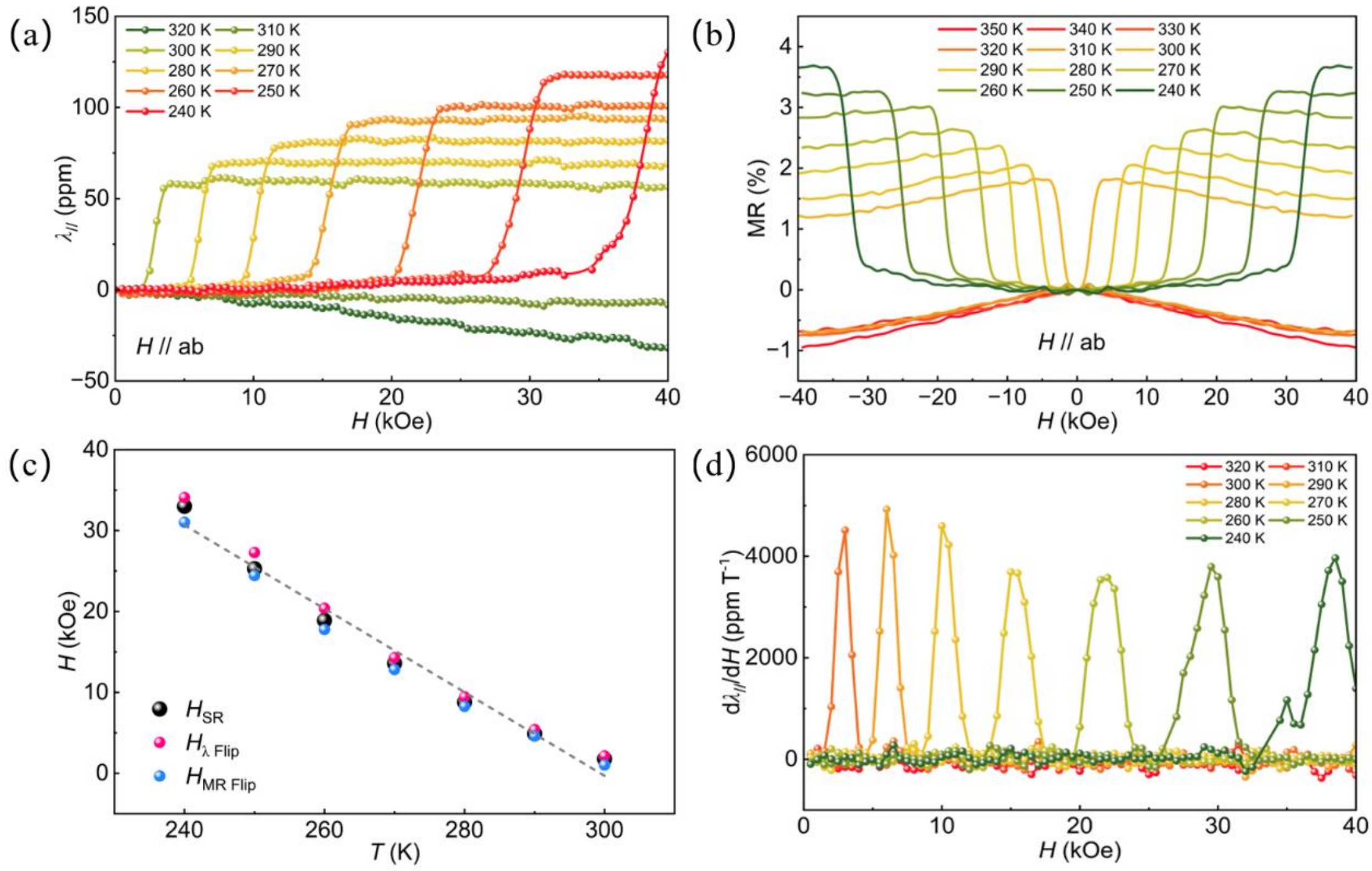

Figure 3. (a) Schematic diagrams illustrating the measurements of in-plane magnetostriction. (b) Magnetic-field dependent magnetostriction at various temperatures. (c) Magnetic-field dependent of the slope of the magnetostriction: $d\lambda/dH$ at various temperatures (d) Temperature dependence of $H_{SR}$ and $H_{\lambda Flip}$.

Interestingly, it is found that the abrupt magnetostrictive jump occurs at a field that closely coincides with the magnetic anomaly observed in the magnetization measurements (Fig. 3(c)). This correlation suggests that the evolution of magnetostrictive behavior is closely associated with the field-induced spin reorientation. Another notable feature is the temperature dependence of the magnetostrictive amplitude. With decreasing temperature, both the critical magnetic field for abrupt magnetostrictive enhancement and the saturated magnetostriction increase monotonously. At 240 K, the magnetostriction exceeds 120 ppm, which is almost comparable to that of convention Fe-based magnetostrictive alloys [7, 25]. Meanwhile, the increase in the critical field is consistent with the strengthening of magnetocrystalline anisotropy upon cooling. This correlation suggests that the behavior of magnetostriction in magnetic topological materials is governed not only by the intrinsic spin-lattice coupling [26], but also, and perhaps more importantly by, the nature where the spin configuration evolves under an external magnetic field. Notably, the critical fields of the magnetostriction jump also closely coincide with those identified from the magneto-transport measurement (Fig. 3(b)), providing an additional indication of the correspondence between the field-induced spin reorientation and the abrupt magnetostrictive response.

Beyond the magnetostriction, the differential magnetostrictive response, defined as $d\lambda/dH$, is extracted from this data and is plotted as a function of temperature in Fig.3 (d). Through the Maxwell relation $(d\lambda/dH)_{\sigma}=(dM/d\sigma)_{H}$, the magnetostriction response is thermodynamically linked to the reciprocal piezomagnetic coefficient, which is favorable to further evaluate the magnetostrictive properties [27]. Notably, the field-induced magnetostrictive jump in $TbMn_6Sn_6$ is accompanied by a pronounced peak in $d\lambda/dH$, reaching approximately 4000 ppm $T^{-1}$, with a maximum exceeding 5000 ppm $T^{-1}$. Although this value is lower than that for Terfenol-D, for whom respectively with 13000 ppm $T^{-1}$ [28], it substantially exceeds directionally solidified Galfernol-D with 1700 ppm $T^{-1}$ [29], and topological $Mn_3Sn$ in earlier report with approximately 10 and 15 ppm $T^{-1}$ [17]. This result suggests the potential for magnetomechanical conversion in magnetic topological materials through tailoring spin configuration.

The temperature dependence of in-plane thermal expansion coefficient was also measured under different in-plane magnetic fields (Fig. 3(a)). The thermal expansion coefficient is positive upon the whole temperature range. While pronounced kinks were observed on each curve, for whom the corresponding temperature consistent with the field-induced spin-reorientation identified from magnetization measurements as well. Such results demonstrates that the SRT in $TbMn_6Sn_6$ is accompanied by a substantial lattice response.

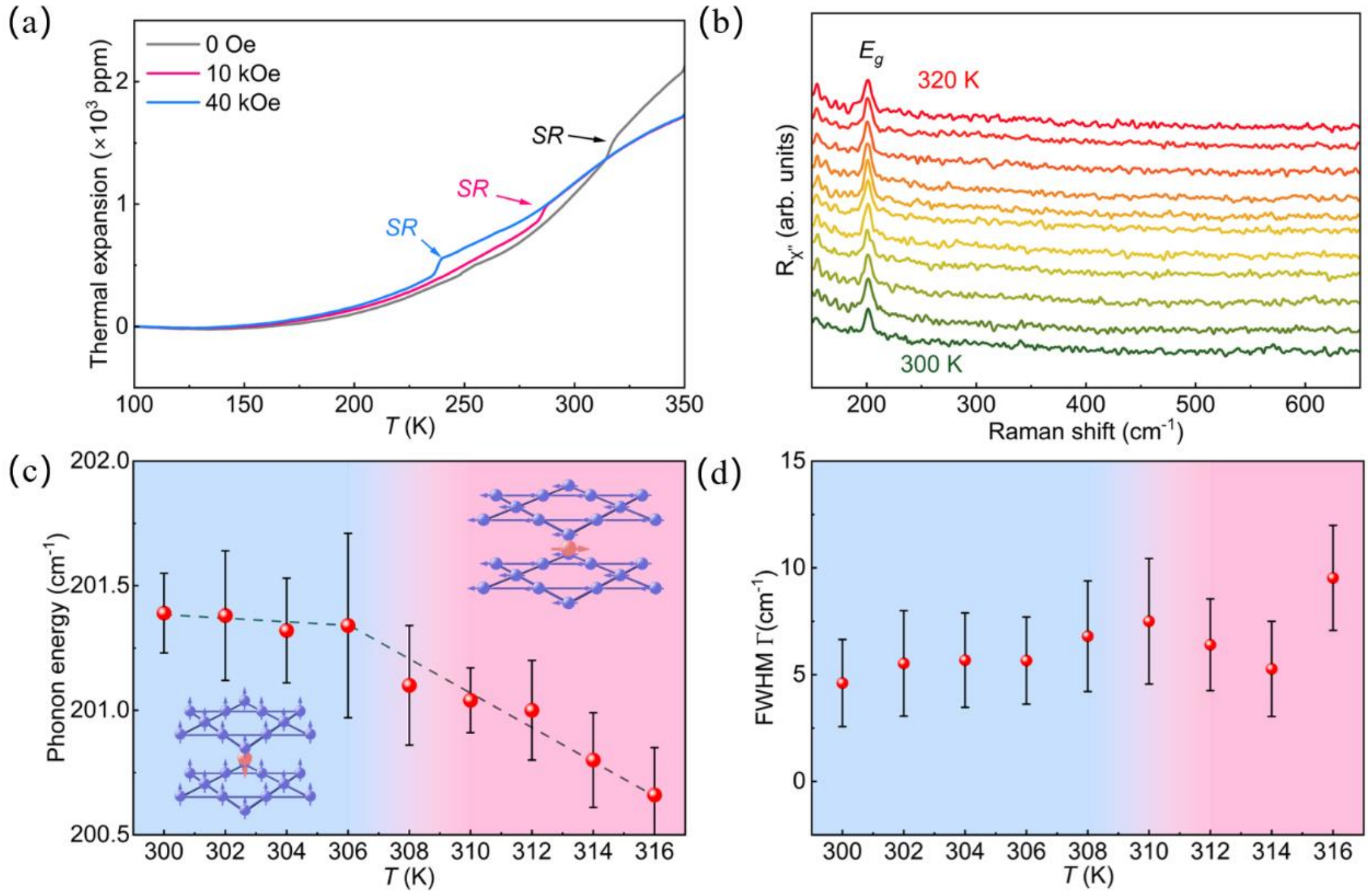

Figure 4. (a) Temperature dependence of thermal expansion under various in-plane magnetic field. (b) Raman spectral recorded at various temperatures across $T_{SR}$. Energy and linewidth temperature dependence of $E_g$ mode in $TbMn_6Sn_6$ (c, d) determined from (b). The insets in (c) schematically illustrate the predominantly out-of-plane and in-plane spin configurations on either side of $T_{SR}$, highlighting the possible spin-orientation-dependent lattice distortion associated with the spin-reorientation transition.

To further investigate the evolution of the lattice dynamic upon SRT, temperature-dependent Raman spectroscopy was employed to investigate the lattice-dynamical response across the $T_{SR}$ (Fig. 4(b)). As shown in Fig. 4(c), the in-plane $E_g$ mode phonon softens progressively upon warming, consistent with the combined effects of thermal expansion and anharmonic phonon interactions. Notably, the phonon energy exhibits an apparent change in slope around $T_{SR}$, suggesting an additional phonon renormalization, where the effective restoring forces of the in-plane vibrational mode modifies as the magnetic state changes. The structural insets in Fig. 4(c) schematically illustrate how the different spin configurations may be associated with distinct equilibrium lattice strains, providing a possible magnetoelastic link between the spin reorientation and the observed thermal expansion anomalies, as well as the magnetostriction. In addition, the phonon linewidth generally broadens with increasing temperature but exhibits a slight narrowing near $T_{SR}$, followed by renewed broadening at higher temperatures (Fig. 4(d)). This non-monotonic behavior suggests a modification of phonon damping or scattering processes around the $T_{SR}$. Although the linewidth anomaly does not uniquely identify the underlying scattering mechanism, its occurrence in the same temperature as the phonon-energy anomaly provides evidence that the lattice-dynamical response is sensitive to the magnetic transition.

The magnetostrictive response of $TbMn_6Sn_6$ can be understood in terms of the interplay between competing magnetic anisotropies and magnetoelastic coupling. The competition between the Tb-associated uniaxial anisotropy and the easy-plane anisotropy of the Mn sublattice, mediated by Tb-Mn exchange, governs the stability of different spin configurations [20, 30]. Below $T_{SR}$, an in-plane magnetic field can drive a further reorientation of the magnetic moments by modifying the relative energies of these configurations. Because the magnetoelastic energy depends on the spin configuration, this reorientation can shift the equilibrium lattice strain, providing a physical basis for the abrupt magnetostrictive response. The correspondence between the characteristic fields of the magnetization anomaly and the enhanced magnetostriction supports their common association with field-induced spin reorientation. Meanwhile, the thermal-expansion response and the anomalies in the temperature dependence of the in-plane $E_g$ mode phonon frequency and linewidth indicate that the lattice response and vibrational dynamics are also sensitive to the SRT. Taken together, our results establish a close phenomenological connection between spin reorientation, magnetoelastic strain, and phonon dynamics in $TbMn_6Sn_6$, highlighting spin-orientation-dependent magnetoelastic coupling as a useful strategy for exploring high-sensitive magntostrictive materials.

## Conclusion

In summary, we have investigated the magnetostrictive response of $TbMn_6Sn_6$ across its SRT and established a close association between field-induced spin reorientation and the enhanced strain response. Below $T_{SR}$, the application of an in-plane magnetic field produces an abrupt increase in magnetostriction, accompanied by a pronounced piezomagnetic response. The correspondence between the critical fields of the magnetostriction and magnetization anomalies, together with the related magneto-transport features, supports their association with field-induced spin reorientation. In addition, thermal-expansion measurements and temperature-dependent Raman spectroscopy reveal changes in the lattice response and $E_g$ mode phonon dynamics around $T_{SR}$, providing complementary evidence of the coupling between magnetic reconfiguration and lattice degrees of freedom. These observations are consistent with a scenario in which competing magnetic anisotropies of the Tb and Mn sublattices, mediated by Tb-Mn exchange, stabilize distinct spin configurations with different equilibrium strains. Our results highlight spin-orientation-dependent magnetoelastic coupling as a useful framework for understanding the enhanced field-induced strain response in $TbMn_6Sn_6$.

**Acknowledgments**

We acknowledge financial support from the National Natural Science Foundation of China (NSFC 52471214, 52271016), Ningbo Yongjiang Talent Introduction Programme (2022A-090-G), the Max Planck Partner Group program, the Zhejiang Provincial Natural Science Foundation of China (Grant No. LQN26E010015)

**Conflicts of Interest**

The authors declare no conflict of interest.